\documentclass[onecolumn,amsmath,showpacs,nofootinbib,12pt]{revtex4-1}
\usepackage{graphicx}
\usepackage{dcolumn}
\usepackage{bm}
\usepackage{color} 
\usepackage{slashed}
\begin{document}
\newcommand{\hs}{\hspace*{0.5cm}}
\newcommand{\vs}{\vspace*{0.5cm}}
\newcommand{\be}{\begin{equation}}
\newcommand{\ee}{\end{equation}}
\newcommand{\bea}{\begin{eqnarray}}
\newcommand{\eea}{\end{eqnarray}}
\newcommand{\ben}{\begin{enumerate}}
\newcommand{\een}{\end{enumerate}}
\newcommand{\bde}{\begin{widetext}}
\newcommand{\ede}{\end{widetext}}
\newcommand{\nn}{\nonumber}
\newcommand{\crn}{\nonumber \\}
\newcommand{\Tr}{\mathrm{Tr}}
\newcommand{\non}{\nonumber}
\newcommand{\noi}{\noindent}
\newcommand{\al}{\alpha}
\newcommand{\la}{\lambda}
\newcommand{\bet}{\beta}
\newcommand{\ga}{\gamma}
\newcommand{\va}{\varphi}
\newcommand{\om}{\omega}
\newcommand{\pa}{\partial}
\newcommand{\+}{\dagger}
\newcommand{\fr}{\frac}
\newcommand{\bc}{\begin{center}}
\newcommand{\ec}{\end{center}}
\newcommand{\Ga}{\Gamma}
\newcommand{\de}{\delta}
\newcommand{\De}{\Delta}
\newcommand{\ep}{\epsilon}
\newcommand{\varep}{\varepsilon}
\newcommand{\ka}{\kappa}
\newcommand{\La}{\Lambda}
\newcommand{\si}{\sigma}
\newcommand{\Si}{\Sigma}
\newcommand{\ta}{\tau}
\newcommand{\up}{\upsilon}
\newcommand{\Up}{\Upsilon}
\newcommand{\ze}{\zeta}
\newcommand{\ps}{\psi}
\newcommand{\Ps}{\Psi}
\newcommand{\ph}{\phi}
\newcommand{\vph}{\varphi}
\newcommand{\Ph}{\Phi}
\newcommand{\Om}{\Omega}
\newcommand{\AdrHEPC}{Phenikaa Institute for Advanced Study, Phenikaa University, Yen Nghia, Ha Dong, Hanoi 100000, Vietnam}

\title{Scotoseesaw mechanism from a $Z_3$ symmetry of matter} 

\author{Doan Minh Luong}
\email{21010345@st.phenikaa-uni.edu.vn}
\author{Phung Van Dong} 
\email{dong.phungvan@phenikaa-uni.edu.vn (corresponding author)}
\affiliation{\AdrHEPC} 

\date{\today}

\begin{abstract}

We show that the neutrino mass generation and the dark matter stability can be governed by the center of the QCD group, which is a $Z_3$ group. Three right-handed neutrinos $N_{1,2,3R}$ transform under $Z_3$ as $1,w,w^2$, where $w=e^{i2\pi/3}$ is the cube root of unity, and they couple to usual lepton doublets via the usual Higgs doublet $H$ and two new scalar doublets $\eta,\chi$, which transform under $Z_3$ as $1,w^2,w$, respectively. This leads to a scotoseesaw mechanism in which the seesaw and scotogenic neutrino mass generations are induced by the Majorana $N_{1R}$ mass and the Dirac $N_{2,3R}$ mass, respectively. Although the lightest of the $Z_3$ fields is stabilized, responsible for dark matter, the model lacks an explanation for relic density and/or direct detection. The issue can be solved in a $U(1)_{B-L}$ gauge completion of the model, for which the center of the QCD group is isomorphic to $Z_3=\{1,T,T^2\}$ for $T=w^{3(B-L)}$.

\end{abstract} 

\maketitle

\section{Introduction} 

Of the exact symmetries in physics, the symmetry of the baryon-minus-lepton number, say $B-L$, causes curiosity. There is no necessary principle for the $B-L$ symmetry, since it directly results from the standard model Lagrangian. Indeed, every interaction of the standard model separately preserves $B$ and $L$ so that $B-L$ is conserved and anomaly-free, given that right-handed neutrinos are simply included. 

The evidence of neutrino oscillations \cite{kajita,mcdonald} acquires that neutrinos have nonzero masses, which break $B-L$ if they are Majorana particles. Additionally, the observation of cosmological baryon asymmetry \cite{planck} requires that $B-L$ must be broken too. If $B-L$ is broken by two units, it is reduced to a conserved symmetry $Z_6=Z_2\otimes Z_3$ \cite{dong1}. If $B-L$ is broken by one unit, it is reduced to a conserved symmetry $Z_3$ \cite{dong2}. As a matter of fact, every $B-L$ breaking would leave a common residual $Z_3$ group conserved, composed of $\{1,T,T^2\}$, where $T=w^{3(B-L)}$ obeys $T^3=1$, and $w=e^{i 2\pi/3}$ is the cube root of unity. 

Indeed, this $Z_3$ symmetry, which transforms nontrivially only for quarks, namely $T=w$ for $q$, $T=w^2$ for $q^c$, and $T=1$ for other fields (leptons, Higgs fields, and gauge fields), is accidentally conserved by the $SU(3)_C$ color symmetry, where the $Z_3$ group is isomorphic to the center of $SU(3)_C$. Since the $SU(3)_C$ gauge symmetry is always conserved and unbroken, the $Z_3$ symmetry is absolutely preserved, in contrast to the matter parity $P_M=(-1)^{3(B-L)+2s}$, as in the $Z_2$ group above, which is possibly broken \cite{martin}. 

Further, if a dark matter candidate \cite{bertone,arcadi}, such as a right-handed neutrino, is charged under the $Z_3$ group, it cannot decay to quarks, in spite of the fact that both transform nontrivially under $Z_3$ and if the dark matter is heavier than quarks. Indeed, since the dark matter candidate is color neutral, it cannot decay to a single quark, because of color conservation. If the dark matter candidate decays to several quarks, the QCD symmetry demands that the final state must compose $qq^c$ and/or $qqq$ due to color conservation but is trivial under $Z_3$, which is suppressed by $Z_3$. Hence, the dark matter candidate is absolutely stabilized, due to the color symmetry, in addition to $Z_3$.  

We would like to suggest in this work that three right-handed neutrinos ($N_{1,2,3R}$) exist as irreducible representations of the $Z_3$ group, i.e. transforming (almost nontrivially) according to the center of the color group, such as $T=1,w,w^2$ assigned to $N_{1,2,3R}$, respectively. That said, $N_{2,3R}$ reveals a Dirac fermion state under $Z_3$, i.e. $N_{2L}\equiv N^c_{3R}$ so that the Dirac fermion mass $D_2 \bar{N}_{2L} N_{2R}+H.c.$ is invariant under $Z_3$. Additionally, $N_{1R}$ is a Majorana field with mass $\fr 1 2 M_1 \bar{N}^c_{1R} N_{1R}+H.c.$ invariant under $Z_3$ by itself. This implies that neutrino masses might be generated by a scotoseesaw mechanism \cite{ss1,ss2}, where the seesaw part \cite{seesaw1,seesaw2,seesaw3,seesaw4,seesaw5} is governed by $N_{1R}$, which is trivial under $Z_3$, coupled to neutrinos and the Higgs boson, while the scotogenic part \cite{tao,ma} is set by $N_{2L,R}$, which is nontrivial under $Z_3$, coupled to neutrinos and appropriate $Z_3$ scalar doublets. This proposal predicts a unique Dirac dark matter candidate $N_2$, besides potential $Z_3$ scalar candidates. 

The scotoseesaw has been extensively studied in \cite{ss3,ss4,ss5,ss6,ss7,ss8,dongsg,ss9,ss10,ss11}, in which the dark sector symmetry is centrally discussed. The novelty of this work is that the dark matter stability may be due to the center of the color group for which the dark matter is charged. However, this mechanism lacks an explanation for dark matter relic density and/or direct detection. This problem is resolved if one supplies a $U(1)_{B-L}$ gauge completion for $Z_3$.  

In what follows, we first present the toy model based only on the center of the QCD group in detail in Sec. \ref{z3ss}. In this case, we identity the scalar mass spectrum and investigate the neutrino mass generation. We then give comments on dark matter stability and detection. We next present a gauge completion for the toy model in detail in Sec. \ref{gc}. In this gauge completion, we reexamine the neutrino mass generation and investigate the relevant dark matter density and detection phenomena. Finally, we conclude this work in Sec. \ref{concl}.                      

\section{\label{z3ss} The toy model}

The QCD sector supplies a center of $SU(3)_C$, well-established as the $Z_3$ group, which transforms nontrivially on quarks. What happens if this $Z_3$ symmetry governs the exotic neutrino sector?

\subsection{\label{pro1} Proposal}

\begin{table}[h]
\bc
\begin{tabular}{lccccc}
\hline\hline 
Field & $SU(3)_C$ & $SU(2)_L$ & $U(1)_Y$ & $Z_3$\\
\hline
$l_{aL} = \begin{pmatrix}
\nu_{aL}\\
e_{aL}\end{pmatrix}$ & 1 & 2 & $-1/2$ & $1$\\
$e_{aR}$ & 1 & 1 & $-1$ & $1$\\
$q_{aL}= \begin{pmatrix}
u_{aL}\\
d_{aL}\end{pmatrix}$ & 3 & 2 & $1/6$ & $w$\\
$u_{aR}$ & 3 & 1 & $2/3$ & $w$\\
$d_{aR}$ & 3 & 1 & $-1/3$ & $w$ \\
$N_{1R}$ & 1 & 1 & 0 & $1$\\
$N_{2R}$ & 1 & 1 & 0 & $w$\\
$N_{3R}$ & 1 & 1 & 0 & $w^2$\\
$H=\begin{pmatrix}
H^+\\
H^0\end{pmatrix}$ & 1 & 2 & $1/2$ & 1 \\  
$\eta =\begin{pmatrix}
\eta^0\\
\eta^-\end{pmatrix}$ & 1 & 2 & $-1/2$ & $w^2$\\  
$\chi =\begin{pmatrix}
\chi^0\\
\chi^-\end{pmatrix}$ & 1 & 2 & $-1/2$ & $w$\\  
\hline\hline 
\end{tabular}
\caption[]{\label{tab1} Field representation content.}
\ec
\end{table}    

We add three right-handed neutrinos $N_{1,2,3R}$, which transform under the center of the color group $Z_3$ as $1,w,w^2$ for $w=e^{i2\pi/3}$, respectively, despite the fact that $N_{1,2,3R}$ are color neutral. Notice that the usual leptons $l_a=(\nu_a,e_a)$ are trivial under $Z_3$, while the quarks $q_a=(u_a,d_a)$ transform as $w$ under $Z_3$, where $a=1,2,3$ is a family index. Besides the usual Higgs doublet $H$, we introduce two $Z_3$ scalar doublets $\eta,\chi$, which transform as $w^2,w$ under $Z_3$ and couple $N_{2,3R}$ to the lepton doublets, respectively. The field representation content of the model is collected in Table \ref{tab1}.

The total Lagrangian takes the form,
\be \mathcal{L}=\mathcal{L}_{\mathrm{kinetic}}+\mathcal{L}_{\mathrm{Yukawa}}-V_{\mathrm{scalar}},\ee where the first part contains kinetic terms and gauge interactions. The Yukawa part is 
\bea \mathcal{L}_{\mathrm{Yukawa}} &=& h^u_{ab} \bar{q}_{aL}\tilde{H}u_{bR}+ h^d_{ab} \bar{q}_{aL} H d_{bR}+h^e_{ab}\bar{l}_{aL}He_{bR}+\crn
&&+ h^\nu_{a1}\bar{l}_{aL}\tilde{H}N_{1R}+h^\nu_{a2}\bar{l}_{aL}\eta N_{2R}+h^\nu_{a3}\bar{l}_{aL}\chi N_{3R}\crn
&& -\fr 1 2 M_{1} \bar{N}^c_{1R}N_{1R} -D_{2}\bar{N}^c_{3R} N_{2R}+H.c.,\eea where $b=1,2,3$ is a family index as $a$ is. Additionally, $h$'s stand for Yukawa couplings, while $M_1$ and $D_2$ denote Majorana and Dirac masses coupled to $N_{1R}$ and $N_{2,3R}$, respectively. 

The scalar potential is 
 \bea V_{\mathrm{scalar}} &=& \mu^2_1 H^\dagger H  + \mu^2_2 \eta^\dagger \eta + \mu^2_3 \chi^\dagger \chi \crn
 && + \la_1(H^\dagger H)^2 +\la_2(\eta^\dagger \eta)^2+\la_3(\chi^\dagger \chi)^2\crn
 &&+\la_4 (H^\dagger H)(\eta^\dagger \eta) +\la_5 (H^\dagger H)(\chi^\dagger \chi)+\la_6(\eta^\dagger \eta)(\chi^\dagger \chi)\crn
 && +\la_7 (H^\dagger \eta)(\eta^\dagger H)+\la_8 (H^\dagger \chi)(\chi^\dagger H)+\la_9(\eta^\dagger \chi)(\chi^\dagger \eta)\crn
 &&+\left[\la_{10} (H\eta)(H\chi) +H.c.\right],\label{vhec}\eea where $\la$'s stand for dimensionless couplings, while $\mu$'s have a mass dimension. The coupling $\la_{10}$ is assumed to be real, since its phase if present is removed by redefining appropriate scalar fields. The desirable vacuum structure demands that \be \mu^2_{1}<0\ \mathrm{and}\ \mu^2_{2,3}>0.\label{dt101e}\ee Further, the scalar potential bounded below requires 
 \bea &&\la_{1,2,3}>0,\label{dt101a}\\
 && \la_4+\la_7\Theta(-\la_7)>-2\sqrt{\la_1\la_2},\label{dt101b}\\
 && \la_5+\la_8\Theta(-\la_8)>-2\sqrt{\la_1\la_3},\label{dt101c}\\
 && \la_6+\la_9\Theta(-\la_9)>-2\sqrt{\la_2\la_3},\label{dt101d}\eea where $\Theta(x)$ is the Heaviside step function. There still exist complicated conditions in order to ensure the quartic coupling matrix to be copositive responsible for the vacuum stability, which can be extracted with the aid of \cite{kannikeadd1}. The extra conditions must be presented for the potential to be perturbative and unitarity but is skipped for brevity.     

\subsection{\label{scalar} Scalar spectrum}

The $Z_3$ conservation demands that $\eta,\chi$ have vanished vacuum expectation values (VEVs), while $H$ can develop a nonzero VEV. Additionally, $H$ does not mix with $\eta,\chi$. Expanding $H^0=(v+h+i G_Z)/\sqrt{2}$, the doublet $H$ is given by  
\be H=\begin{pmatrix}
G^+_W\\
\fr{1}{\sqrt{2}}(v+h+i G_Z)
\end{pmatrix}, \ee where $v=\sqrt{-\mu^2_1/\la_1}=246$ GeV is the weak scale, while $G^+_W = H^+$ and $G_Z$ are massless Goldstone bosons eaten by the gauge bosons $W^+$ and $Z$, respectively. The field $h$ is the usual Higgs boson with mass $m_h=\sqrt{2\la_1} v$.

The charged $Z_3$ scalar fields $\eta^-$ and $\chi^-$ are physical mass eigenstates by themselves, whose masses are given, respectively, by
\bea m^2_{\eta^-} &=& \mu^2_2+\fr 1 2 (\la_4+\la_7) v^2,\\ 
m^2_{\chi^-} &=& \mu^2_3+\fr 1 2 (\la_5+\la_8) v^2. \eea By contrast, the neutral $Z_3$ scalar fields $\eta^0$ and $\chi^{0}$ can be expanded as follows \be \eta^0=\fr{R_\eta+iI_\eta}{\sqrt{2}},\hs \chi^0=\fr{R_\chi+i I_\chi}{\sqrt{2}},\ee and their real (CP-even) and imaginary (CP-odd) fields mix via mass matrices,
\bea V_{\mathrm{scalar}} &\supset& \fr 1 2 \begin{pmatrix} 
R_\eta & R_\chi \end{pmatrix}
\begin{pmatrix} \mu^2_2+\fr{\la_4}{2}v^2 & \fr{\la_{10}}{2}v^2\\
\fr{\la_{10}}{2}v^2& \mu^2_3 +\fr{\la_5}{2}v^2
\end{pmatrix}
\begin{pmatrix} 
R_\eta\\
R_\chi
\end{pmatrix}\crn
&& +\fr 1 2 \begin{pmatrix} 
I_\eta & I_\chi \end{pmatrix}
\begin{pmatrix} \mu^2_2+\fr{\la_4}{2}v^2 & -\fr{\la_{10}}{2}v^2\\
-\fr{\la_{10}}{2}v^2& \mu^2_3 +\fr{\la_5}{2}v^2
\end{pmatrix}
\begin{pmatrix} 
I_\eta\\
I_\chi
\end{pmatrix}.\label{ddtt001} \eea

To diagonalize the neutral $Z_3$ scalar sector, we define a mixing angle $\theta$, such that 
\be t_{2\theta}=\fr{\la_{10} v^2}{\mu^2_3-\mu^2_2+\fr 1 2 (\la_5-\la_4)v^2}. \ee The mass matrix of the real fields in (\ref{ddtt001}) yields physical mass eigenstates to be \be R_1 = c_\theta R_\eta -s_\theta R_\chi,\hs R_2 = s_\theta R_\eta +c_\theta R_\chi, \ee with respective mass eigenvalues, 
\bea m^2_{R_1} &=&\fr 1 2 \left\{\mu^2_2+\mu^2_3+ (\la_4+\la_5)v^2/2-\sqrt{\left[\mu^2_2-\mu^2_3+ (\la_4-\la_5)v^2/2\right]^2+\la^2_{10}v^4}\right\},\\
m^2_{R_2} &=&\fr 1 2 \left\{\mu^2_2+\mu^2_3+ (\la_4+\la_5)v^2/2+\sqrt{\left[\mu^2_2-\mu^2_3+ (\la_4-\la_5)v^2/2\right]^2+\la^2_{10}v^4}\right\}.\eea Whilst, the mass matrix of the imaginary fields in (\ref{ddtt001}) supplies physical mass eigenstates, 
\be I_1 = c_\theta I_\eta + s_\theta I_\chi,\hs I_2 = -s_\theta I_\eta +c_\theta I_\chi, \ee with respective mass eigenvalues, 
\bea m^2_{I_1} &=&\fr 1 2 \left\{\mu^2_2+\mu^2_3+ (\la_4+\la_5)v^2/2-\sqrt{\left[\mu^2_2-\mu^2_3+ (\la_4-\la_5)v^2/2\right]^2+\la^2_{10}v^4}\right\},\\
m^2_{I_2} &=&\fr 1 2 \left\{\mu^2_2+\mu^2_3+ (\la_4+\la_5)v^2/2+\sqrt{\left[\mu^2_2-\mu^2_3+ (\la_4-\la_5)v^2/2\right]^2+\la^2_{10}v^4}\right\}.\eea

It is clear that $R_1$ and $I_1$ have the same mass which is identified as a physical neutral complex field, such as 
\be H_1 \equiv \fr{R_1+i I_1}{\sqrt{2}} = c_\theta \eta^0 - s_\theta \chi^{0*},\ee with mass $m_{H_1}=m_{R_1}=m_{I_1}$. Similarly, $R_2$ and $I_2$ possess the same mass which is identified as a physical neutral complex field, 
\be H_2\equiv \fr{R_2-i I_2}{\sqrt{2}} = s_\theta \eta^0 + c_\theta \chi^{0*},\label{ddtt002}\ee with mass $m_{H_2}=m_{R_2}=m_{I_2}$. In this way, the common, but opposite, mixing angle of the real and imaginary parts, i.e. $\theta$, is just the $\eta^0$-$\chi^{0*}$ mixing angle. 

The physical neutral $Z_3$ fields obtain approximate masses,    
\bea m^2_{H_1}
&\simeq & \mu^2_2+\fr{\la_4}{2}v^2-\fr{\fr 1 4 \la^2_{10}v^4}{\mu^2_{3}-\mu^2_2+\fr 1 2 (\la_5-\la_4)v^2},\\
m^2_{H_2}&\simeq & \mu^2_3+\fr{\la_5}{2}v^2+\fr{\fr 1 4 \la^2_{10}v^4}{\mu^2_{3}-\mu^2_2+\fr 1 2 (\la_5-\la_4)v^2}.\eea Here the approximation assumes that the $\eta^0$-$\chi^{0*}$ mixing angle is small. Notice that in this case, $H_1\simeq \eta^0$ and $H_2\simeq \chi^{0*}$, which indicates why $H_2$ has a nature of a conjugate field as in~(\ref{ddtt002}). Further, substituting the approximated masses, the mixing angle takes the form,
\be s_{2\theta}\simeq \fr{\la_{10}v^2}{m^2_{H_2}-m^2_{H_1}}.\label{nmxa}\ee  

\subsection{\label{nmass} Neutrino mass generation}  

\begin{figure}[h]
\bc
\includegraphics[scale=1]{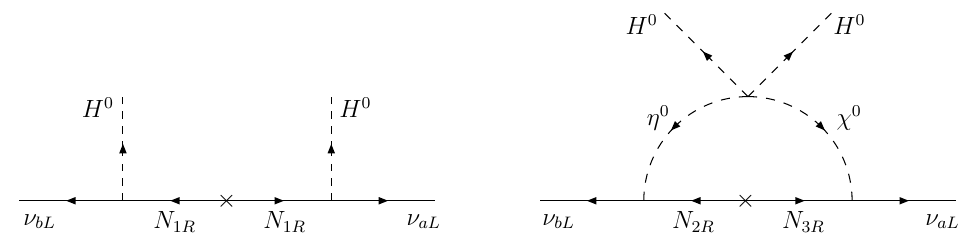}
\caption[]{\label{fig1} Scotoseesaw scheme governed by the $Z_3$ symmetry of matter.}
\ec
\end{figure}

The neutrino mass generation is derived by a scotoseesaw mechanism, as given in diagrams in Fig. \ref{fig1}. In the mass basis, the loop diagram is mediated by a Dirac fermion $N_{2L,R}$ and two scalars $H_{1,2}$ instead. It is straightforward to derive the neutrino mass
\be m_\nu = m^{\mathrm{tree}}_\nu + m^{\mathrm{loop}}_\nu, \ee where 
\be (m^{\mathrm{tree}}_\nu)_{ab}  = -\fr{h^\nu_{a1}h^\nu_{b1}v^2}{2M_1},\label{nmtree}\ee
and 
\be (m^{\mathrm{loop}}_\nu)_{ab} = \fr{c_\theta s_\theta (h^\nu_{a2} h^\nu_{b3}+h^\nu_{a3}h^\nu_{b2}) D_2}{32\pi^2}\left(\fr{m^2_{H_2}\ln \fr{D^2_2}{m^2_{H_2}}}{D^2_2-m^2_{H_2}}-\fr{m^2_{H_1}\ln \fr{D^2_2}{m^2_{H_1}}}{D^2_2-m^2_{H_1}}\right).\ee 

We demand that the Dirac fermion $N_{2L,R}$ is radically lighter than $H_{1,2}$, responsible for dark matter, which implies $D^2_2\ll m^2_{H_{1,2}}$. Additionally, we assume the masses of $N_{1R}$ and $H_{1,2}$ to be at TeV regime, i.e. $M_1\sim m_{H_{1,2}}\sim $ TeV, responsible for neutrino mass. The last proportion leads to $\mu_{2,3}\sim $ TeV. Since $\mu_{2}$ and $\mu_3$ are generically separated, the masses of $H_1$ and $H_2$ are separated too. The mixing angle $\theta$ is small, given by (\ref{nmxa}). Hence, the radiative neutrino mass is approximated as
\be
(m^{\mathrm{loop}}_\nu)_{ab} \simeq \fr{\la_{10}(h^\nu_{a2} h^\nu_{b3}+h^\nu_{a3}h^\nu_{b2})D_2}{64\pi^2}\fr{v^2}{m^2_{H_2}-m^2_{H_1}} \ln \fr{m^2_{H_2}}{m^2_{H_1}}.\label{nmloop}\ee Because $D_2$ is at/below the weak scale, the radiative neutrino mass is suppressed by (i) the loop factor $1/32\pi^2\ll 1$ and (ii) the double seesaw $(v/m_{H_{1,2}})^2\ll 1$.  

Comparing the tree-level neutrino mass (\ref{nmtree}) with the measured value of Atmospheric neutrino mass (large) splitting, i.e. $m^{\mathrm{tree}}_\nu\sim 0.1$ eV, we derive $h^\nu_{a1}\sim 10^{-5}$, like the Yukawa coupling of electrons. Further, taking $\la_{10}\sim 10^{-3}$, $h^\nu_{a2}h^\nu_{b3}+h^\nu_{a3}h^\nu_{b2}\sim 10^{-4}$, and $v/m_{H_{1,2}}\sim 0.1$, comparing the loop-level neutrino mass (\ref{nmloop}) with the measured value of the Solar neutrino mass (small) splitting, i.e. $m^{\mathrm{loop}}_\nu\sim 0.01$ eV, yields $D_2\sim 10$~GeV. 

\subsection{\label{dmatter} Remarks on dark matter}

In the early universe, the Dirac fermion $N_{2L,R}$ with mass $D_2\sim 10$ GeV annihilates to usual leptons via $t$-channel diagrams exchanged by $\eta,\chi$. But the annihilation cross section is too small due to $h^\nu_{a2(3)}\sim 10^{-2}$ to set the correct density. It is noted that $H_{1,2}$ belong to the weak doublets significantly interact with nuclei via $Z$-exchange, which were already ruled out by the direct detection experiment, i.e. they cannot be interpreted as a dark matter candidate \cite{barbieriadd2}. Indeed, if $H_1$ is dark matter, for instance, the spin-independent (SI) scattering cross section of it with a nucleus ($\mathcal{N}$) is given by
\be \sigma^{\mathrm{SI}}_{H_1-\mathcal{N}}=\fr{G^2_{\mathrm{F}}m^2_{r}}{2\pi}\left[N-(1-4s^2_W)Z\right]^2,\ee where $N,Z$ are the numbers of neutrons and protons in $\mathcal{N}$, respectively, and $m_{r}=m_{H_1}m_{\mathcal{N}}/(m_{H_1}+m_{\mathcal{N}})$ is the reduced mass. The resultant cross section per a nucleon is $\sigma^{\mathrm{SI}}_{H_1-p,n}\sim 10^{-38}\ \mathrm{cm}^2$ which is 7--9 orders of magnitude beyond the existing bound for dark matter mass in the range few GeV to few TeV \cite{dd3}. Last, but not least, the Dirac dark matter observables are only set when the gauge completion is available. In this case, the uncharacteristic smallness of $h^\nu_{a1}$ can be solved by an inverse seesaw instead.   

\section{\label{gc} $U(1)_{B-L}$ gauge completion: Improved phenomena}

$T=w^{3(B-L)}$ transforms nontrivially only for quarks, hence isomorphic to the center of the color group \cite{dong1,dong2}. We will realize it in a $B-L$ gauge completion of the standard model, which gives rise to the scotoseesaw mechanism, responsible for neutrino mass generation and dark matter stability.   

\subsection{\label{rmodel} $Z_3$ as a residual $U(1)_{B-L}$ symmetry}

As mentioned from the outset, we study a gauge completion by introducing a $U(1)_{B-L}$ gauge symmetry and assigning its charge $X\equiv B-L$ to various right-handed neutrinos $N_{1,2,3,\cdots,n R}$ as $X_{1,2,3,\cdots,n}$, respectively, where we do not (since there is no reason why) limit the number of right-handed neutrinos, $n$. Notice that the usual fermions obtain $X$ charge to be $X=-1$ for leptons and $X=1/3$ for quarks, as usual. The cancellation of the relevant anomalies $[\mathrm{gravity}]^2U(1)_X$ and $[U(1)_X]^3$ demands that 
\bea && X_1+X_2+X_3+\cdots+X_n=-3,\\
&& X^3_1+X^3_2+X^3_3+\cdots+X^3_n=-3,\eea respectively. There is no solution for $n<3$. For $n=3$, the equations reveal integer solutions $X_{1,2,3}=-1,-1,-1$, as usual, and $X_{1,2,3}=-4,-4,+5$, as in \cite{pt}. Unfortunately, the right-handed neutrinos $N_{1,2,3R}$ with such charges transform trivially under $Z_3=\{1,T,T^2\}$ for $T=w^{3(B-L)}$ isomorphic to the center of the color group \cite{dong2}. Alternatively, the equations yield a non-integer solution $X_{1,2,3}=6,-17/3,-10/3$ for right-handed neutrinos $N_{1,2,3R}$ \cite{ebln}, which now transform under $Z_3$ as $1,w,w^2$, respectively. However, this solution does not reveal any simple scotoseesaw mechanism as given above. 

\begin{table}[h]
\bc
\begin{tabular}{lcccccc}
\hline\hline 
Fields & $N_{1R}$ & $N_{2R}$ & $N_{3R}$ & $N_{4R}$ & $N_{5R}$ & $N_{6R}$\\
$X=B-L$ & $-1$ & $-2/3$ & $-1/3$ & $1$ &  $-4/3$ &  $-2/3$\\
 $T=w^{3(B-L)}$ & 1 & $w$ & $w^2$ & 1 & $w^2$ & $w$\\
\hline\hline 
\end{tabular}
\caption[]{\label{tab2} A solution for $B-L$ anomaly with six right-handed neutrinos.}
\ec
\end{table}    

Next, we relax $n$ so that $n\leq 6$ and search for $Z_3$-nontrivial, but not large valued, solutions. The simplest of which is $X_1=-1$, $X_2=-2/3$, $X_3=-1/3$, $X_4=1$, $X_5=-4/3$, and $X_6=-2/3$. Correspondingly, the right-handed neutrinos $N_{1,2,3,4,5,6R}$ as coupled to this solution transform under $Z_3$ as $1,w,w^2,1,w^2,w$, respectively. They are all collected in Table \ref{tab2} for brevity. According to the $Z_3$ transformation property, the right-handed neutrinos $N_{1,2,3,4,5,6R}$ are appropriately conjugated in pairs, so that the relevant mass terms, such as $D_1 N_{1R} N_{4R}+H.c.$, $D_2N_{2R} N_{5R}+H.c.$, and $D_3 N_{3R}N_{6R}+H.c.$, are invariant under $Z_3$. By defining $N_{1L}\equiv N^c_{4R}$, $N_{2L}\equiv N^c_{5R}$, and $N_{3L}\equiv N^c_{6R}$, the solution indeed obeys three vectorlike (or Dirac) fermions $N_{1R,L}$, $N_{2R,L}$, and $N_{3R,L}$ under $Z_3$. Exceptionally, the $Z_3$-trivial fields $N_{1,4R}$ might separately obtain Majorana masses $\fr 1 2 M_1 N_{1R}N_{1R}+\fr 1 2 M_4 N_{4R} N_{4R}+H.c.$, which combined with their Dirac mass $D_1$ define two generic Majorana states, composed of $N_{1,4R}$, with respective masses invariant under $Z_3$ too. It is noted that omitting $X_{1,4}$, the remaining solution for $X_{2,3,5,6}$ becomes that studied in \cite{patra,pt}. However, the presence of $X_{1,4}$, exactly $N_{1,4R}$, indicates to a scotoseesaw in the present setup. Additionally, the Dirac mass of $N_{3R,L}$, i.e. $D_3 \bar{N}_{3L}N_{3R}+H.c.$, breaks $U(1)_{B-L}$, by one charge unit, down to the $Z_3$ group, with $T=w^{3(B-L)}$, as expected \cite{dong2}. Hence, the heavy field $N_{3R,L}$ that determines the residual $Z_3$ symmetry might be integrated out, while $N_{1R,L}$ and $N_{2R,L}$ set the scotoseesaw.      

The scalar doublets $\eta$ and $\chi$ have $X=-1/3$ and $1/3$ so that they couple $N_{2R,L}$ to the usual lepton doublets, respectively. Again, $\eta$ and $\chi$ transform under $Z_3$ as $w^2$ and $w$, respectively. The $U(1)_X$ group is broken by two scalar singlets $\phi_1$ and $\phi_2$ possessing $X=1$ and $2$, which necessarily couple $\phi_1 \bar{N}_{3L} N_{3R}$ and $\phi_2 \bar{N}_{2L} N_{2R}$ (as well as $\phi^*_2 N_{4R}N_{4R}$) to make the Dirac (Majorana) masses, respectively. Notice that $\phi_2$ also couples to $N_{1R}N_{1R}$ but does not contribute to the neutrino mass. As mentioned, $\phi_1$ supplies the Dirac mass of $N_{3R,L}$ and determines the residual $Z_3$ group. Whereas, the VEV of $\phi_2$, which breaks $X$ charge by two unit, conserves the residual $Z_3$ symmetry. Indeed, a $U(1)_X$ transformation has the form, $R=e^{i\al (B-L)}$, where $\al $ is a transforming parameter. It must conserve the vacua $R\langle \phi_{1}\rangle = \langle \phi_{1}\rangle$ and $R\langle \phi_{2}\rangle = \langle \phi_{2}\rangle$, leading to $e^{i\al}=1$ and $e^{i2\al}=1$, respectively. These equations obey $\al=k2\pi$ for $k$ integer. Hence, $R=e^{i k 2\pi (B-L)}=w^{3k(B-L)}=\{1, T, T^2\}$ for $T=w^{3(B-L)}$. It is clear that $T^3=1$, thus $R$ is a symmetry of $Z_3$, as expected. We collect in Table \ref{tab3} field representation content, complete gauge symmetry, and its residual $Z_3$ group. 

\begin{table}[h]
\bc
\begin{tabular}{lccccc}
\hline\hline 
Field & $SU(3)_C$ & $SU(2)_L$ & $U(1)_Y$ & $U(1)_{B-L}$  & $Z_3$\\
\hline
$l_{aL} = \begin{pmatrix}
\nu_{aL}\\
e_{aL}\end{pmatrix}$ & 1 & 2 & $-1/2$ & $-1$ & $1$\\
$e_{aR}$ & 1 & 1 & $-1$ & $-1$ & $1$\\
$q_{aL}= \begin{pmatrix}
u_{aL}\\
d_{aL}\end{pmatrix}$ & 3 & 2 & $1/6$ & $1/3$ & $w$\\
$u_{aR}$ & 3 & 1 & $2/3$ & $1/3$ & $w$\\
$d_{aR}$ & 3 & 1 & $-1/3$ & $1/3$ & $w$ \\
$N_{1R}$ & 1 & 1 & 0 & $-1$ & $1$\\
$N_{2R}$ & 1 & 1 & 0 & $-2/3$ & $w$\\
$N_{3R}$ & 1 & 1 & 0 & $-1/3$ & $w^2$\\
$N_{4R}$ & 1 & 1 & 0 & $1$ & $1$\\
$N_{5R}$ & 1 & 1 & 0 & $-4/3$ & $w^2$\\
$N_{6R}$ & 1 & 1 & 0 & $-2/3$ & $w$\\
$H=\begin{pmatrix}
H^+\\
H^0\end{pmatrix}$ & 1 & 2 & $1/2$ & $0$ & 1 \\  
$\eta =\begin{pmatrix}
\eta^0\\
\eta^-\end{pmatrix}$ & 1 & 2 & $-1/2$ & $-1/3$ & $w^2$\\  
$\chi =\begin{pmatrix}
\chi^0\\
\chi^-\end{pmatrix}$ & 1 & 2 & $-1/2$ & $1/3$ & $w$\\  
$\phi_1$ & 1 & 1 & 0 & $1$ & 1\\
$\phi_2$ & 1 & 1 & 0 & 2 & 1\\
\hline\hline 
\end{tabular}
\caption[]{\label{tab3} Field representation content in the gauge completion.}
\ec
\end{table} 

\subsection{\label{ysint} Lagrangian}   
 
The gauge completion would modify the Lagrangian in the toy model. Apart from the kinetic part that simply adds up the kinetic terms of relevant $U(1)_{B-L}$ fields, the Yukawa interactions now become
\bea \mathcal{L}_{\mathrm{Yukawa}} &=& h^u_{ab} \bar{q}_{aL}\tilde{H}u_{bR}+ h^d_{ab} \bar{q}_{aL} H d_{bR}+h^e_{ab}\bar{l}_{aL}He_{bR}+\crn
&&+ h^\nu_{a1}\bar{l}_{aL}\tilde{H}N_{1R}-D_1\bar{N}^c_{1R} N_{4R}+\fr 1 2 f_{1}\phi_2 \bar{N}^c_{1R}N_{1R}+\fr 1 2 f_2 \phi^*_2 \bar{N}^c_{4R} N_{4R}\crn
&&+h^\nu_{a2}\bar{l}_{aL}\eta N_{2R}+h^\nu_{a3}\bar{l}_{aL}\chi N_{5R} + f_{3} \phi_2 \bar{N}^c_{2R} N_{5R}+ f_{4} \phi_1 \bar{N}^c_{3R} N_{6R}+H.c.,\label{adt3001a} \eea where $D_1$ is a mass parameter, as mentioned, while $f$'s are dimensionless couplings, as $h$'s are.\footnote{Generically, there is a combination of $N_{2R}$ and $N_{6R}$ coupled to $N_{5R}$ in the $f_3$ coupling, and there is another combination of $N_{2R}$ and $N_{6R}$ coupled to $N_{3R}$ in the $f_4$ coupling. The latter is heavy and decoupled along with $N_{3R}$ after $\phi_1$ develops the VEV, while the former is light, coupled to $N_{5R}$ to form a Dirac field after $\phi_2$ develops the VEV. As the Yukawa Lagrangian stands, we have worked in a basis (or relabeled) so that $N_{6R}$ is just the heavy combination, while $N_{2R}$ is the light combination. Additionally, the $h^\nu_{a2}$ coupling includes only the light combination, which subsequently leads to the neutrino mass generation. A similar coupling for the heavy combination does not lead to further physical results and is skipped in this work.} It is clear that the quarks and charged leptons gain appropriate masses via interacting with the Higgs field similar to the toy model. However, the mass generation in the neutral fermion sector is significantly generalized, as shown below.   

Additionally, the scalar potential is now given by \be V_{\mathrm{scalar}}=V(H,\eta,\chi)+V(\phi_1,\phi_2),\ee
where $V(H,\eta,\chi)$ is identical to that in the previous model in (\ref{vhec}), i.e. 
\bea V(H,\eta,\chi) &=& \mu^2_1 H^\dagger H  + \mu^2_2 \eta^\dagger \eta + \mu^2_3 \chi^\dagger \chi \crn
 && + \la_1(H^\dagger H)^2 +\la_2(\eta^\dagger \eta)^2+\la_3(\chi^\dagger \chi)^2\crn
 &&+\la_4 (H^\dagger H)(\eta^\dagger \eta) +\la_5 (H^\dagger H)(\chi^\dagger \chi)+\la_6(\eta^\dagger \eta)(\chi^\dagger \chi)\crn
 && +\la_7 (H^\dagger \eta)(\eta^\dagger H)+\la_8 (H^\dagger \chi)(\chi^\dagger H)+\la_9(\eta^\dagger \chi)(\chi^\dagger \eta)\crn
 &&+\left[\la_{10} (H\eta)(H\chi) +H.c.\right],\eea
while $V(\phi_1,\phi_2)$ contains the $\phi_{1,2}$ potential and the cross-terms between $\phi_{1,2}$ and $H,\eta,\chi$, i.e. 
\bea V(\phi_1,\phi_2) &=& \sum_{i=1,2}\left[\kappa^2_i \phi^\dagger_i \phi_i  + x_i (\phi^\dagger_i \phi_i)^2+\phi^\dagger_i \phi_i (y_i H^\dagger H+z_i \eta^\dagger  \eta+t_i \chi^\dagger \chi)\right] \crn
 && +x_3(\phi^\dagger_1 \phi_1) (\phi^\dagger_2 \phi_2) +\left(\kappa_3 \phi^2_1 \phi^*_2 +H.c.\right),\eea where $\kappa$'s have a mass dimension, while the couplings $x$'s, $y$'s, $z$'s, and $t$'s are dimensionless. Additionally, we can redefine the phases of $\phi_{1,2}$, as appropriate, so that $\kappa_3$ is real. The desirable vacuum structure and the potential stability require $\kappa^2_{1,2}<0$ and \bea && x_{1,2}>0,\\
 && x_3>-2\sqrt{x_1 x_2},\\
 && y_i>-2\sqrt{x_i \la_1},\\
 && z_i > -2\sqrt{x_i \la_2},\\
 && t_i>-2\sqrt{x_i \la_3},\eea for $i=1,2$, as well as those conditions, say (\ref{dt101e}), (\ref{dt101a}), (\ref{dt101b}), (\ref{dt101c}), and (\ref{dt101d}), for $V(H,\eta,\chi)$, besides the extra constraints, as remarked in the above model.  

\subsection{Scalar spectrum}
 
The fields $\eta,\chi$ possess vanished VEVs due to $\mu^2_{2,3}>0$ and the fact that the potential is bounded from below, similar to the previous model. However, the remaining scalars can develop VEVs due to $\kappa^2_{1,2},\mu^2_{1}<0$, such as \bea && \phi_1=\fr{\La_1+S_1+iA_1}{\sqrt{2}},\\  && \phi_2=\fr{\La_2+ S_2+i A_2}{\sqrt{2}},\\  && H=\begin{pmatrix}H^+\\ \fr{1}{\sqrt{2}}(v+S+iA)\end{pmatrix}.\eea 
Substituting these fields to the scalar potential, the gauge invariance requires the linear terms in fields vanished, leading to potential minimization conditions, 
\bea && \mu^2_1 +\la_1 v^2 + \fr 1 2 (y_1\La^2_1+y_2\La^2_2)=0,\\
&& \kappa^2_1 + x_1 \La^2_1 + \fr 1 2 (y_1 v^2 + x_3 \La^2_2)+\sqrt{2} \kappa_3 \La_2=0,\\
&&\kappa^2_2 + x_2 \La^2_2 + \fr 1 2 (y_2 v^2 + x_3 \La^2_2) + \kappa_3 \fr{\La^2_1}{\sqrt{2}\La_2}=0.\label{ctgt}\eea To have a phenomenology as expected, we impose $\La_1\gg v\gg \La_2$, which correspondingly requires $|\kappa_1|\gg |\mu_1|\gg |\kappa_2|$, $y_1\ll \la_1$, and $y_2\ll x_2$. It is noted that the condition $\La_2\ll v$ necessarily generates a dark matter mass at the sub-weak scale, while $\La_1\gg v$ necessarily determines the residual $Z_3$ symmetry, making the $U(1)_{B-L}$ gauge/Higgs and $N_3$ fields decoupled. It is appropriate to take $|\kappa_1|\sim \La_1\sim |\mu_{2,3}|$ at TeV scale like the $Z_3$ scalar masses, while $|\kappa_2|\sim \La_2\sim D_2$ at GeV scale like the $N_{2}$ dark matter mass.

Using the conditions of potential minimization, we find that $G_Z = A$ is a massless Goldstone boson in itself, as associated with the gauge boson $Z$, whereas $A_1$ and $A_2$ mix through a mass matrix given in such order as
\be \begin{pmatrix} -2\sqrt{2} \kappa_3\La_2 & \sqrt{2} \kappa_3 \La_1 \\
\sqrt{2}\kappa_3 \La_1 & -  \fr{\kappa_3\La^2_1}{\sqrt{2} \La_2}\end{pmatrix}.\ee Diagonalizing this mass matrix, we find $G_{Z'} \simeq A_1+2\fr{\La_2}{\La_1}A_2\simeq A_1$ to be a massless Goldstone boson, as coupled to the $U(1)_{B-L}$ gauge boson, the so-called $Z'$, while $J \simeq -2\fr{\La_2}{\La_1} A_1 + A_2\simeq A_2$ is a Marojon field with mass $m^2_J\simeq -\kappa_3\La^2_1/\sqrt{2}\La_2\lesssim \La^2_2$, where we assign it to or below the smallest $B-L$ breaking scale. Indeed, this Majoron is a Goldstone boson coupled to an extra charge, alternative to $B-L$, which is conserved by the potential of $\phi_1,\phi_2$ for $\kappa_3=0$. When the extra charge is approximate, the Majoron would develop a small mass, which must be smaller than the lightest breaking scale of the theory. Since otherwise the conservation of the extra charge demands that $\kappa_3=0$. That said, the parameter $\kappa_3$ that measures the approximate symmetry is given by $\kappa_3\lesssim \La^3_2/\La^2_1$, in agreement with the condition (\ref{ctgt}) in which the relevant terms are equivalently contributed. 

The scalars $S_1$, $S_2$, and $S$ mix, given in such order, via a mass matrix
\be \begin{pmatrix} 2x_1 \La^2_1 & x_3\La_1 \La_2 +\sqrt{2} \kappa_3\La_1 & y_1 v \La_1 \\
x_3\La_1 \La_2 +\sqrt{2} \kappa_3\La_1 & 2 x_2 \La^2_2 - \fr{\kappa_3 \La^2_1}{\sqrt{2}\La_2} & y_2 v \La_2 \\
y_1 v \La_1 & y_2 v \La_2 & 2\la_1 v^2
\end{pmatrix}\ee Because of $\La_1\gg v\gg \La_2$, the field $H' = S_1$ is decoupled from $S,S_2$, obtaining a large mass $m_{H'}\simeq \sqrt{2x_1}\La_1$, where we note that their mixing is suppressed by $v/\La_1 \ll 1$. Additionally, this new Higgs has a VEV, $\La_1\simeq \sqrt{-\kappa^2_1/x_1}$, derived from the minimization conditions. That said, it is appropriate to write down \be \phi_1 \simeq \fr{\La_1+H'+i G_{Z'}}{\sqrt{2}},\ee where $H'$ and $G_{Z'}$ are the new Higgs and Goldstone associated with $B-L$ breaking, dominantly induced by the relevant potential $V_{\mathrm{scalar}}\supset \kappa^2_1 \phi^\dagger_1 \phi_1 + x_1 (\phi^\dagger_1 \phi_1)^2$. 

The mass matrix of the lighter fields $S_2,S$ is determined by the seesaw formula to be 
\be \begin{pmatrix} (2 x_2-\fr{x^2_3}{2x_1}) \La^2_2 - \fr{\kappa_3 \La^2_1}{\sqrt{2}\La_2} & (y_2-\fr{x_3 y_1}{2x_1}) v \La_2 \\
 (y_2-\fr{x_3 y_1}{2x_1}) v \La_2 & (2\la_1 -\fr{y^2_1 }{2x_1})v^2
\end{pmatrix}\ee The standard model Higgs boson $h=c_\xi S-s_\xi S_2$ and the new Higgs boson $h'=s_\xi  S + c_\xi S_2$ are defined via a $S$-$S_2$ mixing angle,
\be t_{2\xi} \simeq -\fr{(y_2-x_3y_1/2x_1)\La_2}{(\la_1-y^2_1/4x_1)v}\ll 1,\ee due to $ \La_2\ll v$, as well as $y_{1,2}\ll \la_1$ and $x_1\sim 1$, as aforementioned. The Higgs masses are approximated by the seesaw formula to be 
\bea m^2_h &\simeq&\left(2\la_1-\fr{y_1^2}{2x_1}\right) v^2,\\
m^2_{h'} &\simeq& \left[\left(2x_2-\fr{x^2_3}{2 x_1}\right)-\fr{(2x_1y_2-x_3y_1)^2}{2x_1(4x_1\la_1-y_1^2)}\right] \La^2_2-\fr{\kappa_3 \La^2_1}{\sqrt{2}\La_2}.\eea Here, $h'$ has a mass at $\La_2$ scale as $J$ does. Notice that $G^+_W=H^+$ is a massless Goldstone boson coupled to $W^+$ by itself. 
Hence, the scalar multiplets $H$ and $\phi_2$ take the form,
\bea H &\simeq &\begin{pmatrix} 
G^+_W\\
\fr{1}{\sqrt{2}}(v+c_\xi h +s_\xi h' +i G_Z)\end{pmatrix},\\
\phi_2 &\simeq &\fr{1}{\sqrt{2}}(\La_2-s_\xi h+c_\xi h'+iJ).\eea   

Above, we have assumed the mixing between $h$ and $h'$ to be small, consistent with the experiment. This is because the $h$-$h'$ mixing would modify the well-measured couplings of the standard model Higgs field, which now bounds $\xi \lesssim 10^{-2}$ \cite{pdg}. Further, the field $h'$ neither significantly couples to fermions nor gives rise to any signal at the LHC. Take, for instance, the Higgs invisible decay to $h',J$, estimated to be $\Ga_h^{\mathrm{inv}}= y^2_2 v^2/16\pi m_h \sim (y_2/10^{-2})^2\times 1$ MeV. This obeys the experiment for $y_2\lesssim 10^{-2}$ \cite{pdg}. It is also noted that the Majoron field $J$ is safe at colliders, since it does not directly couple to fermions. The last comment is that since $\phi_{1,2}$ are singlets, $v=246$ GeV remains unchanged. 

Last, but not least, due to the contribution of $\phi_{1,2}$, the physical charged and neutral $Z_3$ scalar fields and masses may be modified. First, the charged $Z_3$ fields $\eta^-$ and $\chi^-$ are physical fields by themselves as in the toy model, but their masses now become 
 \bea m^2_{\eta^-} &=& \mu^2_2+\fr 1 2 (\la_4+\la_7)v^2+\fr 1 2 z_i \La^2_i,\\
 m^2_{\chi^-} &=& \mu^2_3+\fr 1 2 (\la_5+\la_8)v^2+\fr 1 2 t_i  \La^2_i,\eea respectively, where $i=1,2$ are summed. Second, the neutral $Z_3$ fields whose real and imaginary parts, i.e. $\eta^0=\fr{R_\eta+i I_\eta}{\sqrt{2}}$ and $\chi^0=\fr{R_\chi+i I\chi}{\sqrt{2}}$, now obtain mass matrices,
 \bea V_{\mathrm{scalar}} &\supset& \fr 1 2 \begin{pmatrix} 
R_\eta & R_\chi \end{pmatrix}
\begin{pmatrix} \mu^2_2+\fr{\la_4}{2}v^2 +\fr{z_i}{2} \La^2_i & \fr{\la_{10}}{2}v^2\\
\fr{\la_{10}}{2}v^2& \mu^2_3 +\fr{\la_5}{2}v^2+\fr{t_i}{2} \La^2_i
\end{pmatrix}
\begin{pmatrix} 
R_\eta\\
R_\chi
\end{pmatrix}\crn
&& +\fr 1 2 \begin{pmatrix} 
I_\eta & I_\chi \end{pmatrix}
\begin{pmatrix} \mu^2_2+\fr{\la_4}{2}v^2+\fr{z_i}{2}\La^2_i & -\fr{\la_{10}}{2}v^2\\
-\fr{\la_{10}}{2}v^2& \mu^2_3 +\fr{\la_5}{2}v^2 +\fr{t_i}{2}\La^2_i
\end{pmatrix}
\begin{pmatrix} 
I_\eta\\
I_\chi
\end{pmatrix}. \eea In spite of the $\La_{1,2}$ contribution, these mass matrices of the real and imaginary parts have the same property with those in the toy model, i.e. having the same eigenvalues with opposite mixing angle. That said, the physical neutral $Z_3$ fields, i.e. $H_{1,2}$, can be defined as \be H_1 = c_\theta \eta^0 -s_\theta \chi^{0*},\hs H_2=s_\theta \eta^0+c_\theta \chi^{0*}, \ee which are analogous to those in the toy model, but the $\eta^0$-$\chi^{0*}$ mixing angle is now changed, 
\be t_{2\theta}=\fr{\la_{10} v^2}{\mu^2_3-\mu^2_2+\fr 1 2 (\la_5-\la_4) v^2 +\fr 1 2 (t_i-z_i)\La^2_i}.\ee Additionally, their masses are now given by 
\bea m^2_{H_{1,2}} &=&\fr 1 2 \left\{\mu^2_2+\mu^2_3+ (\la_4+\la_5)v^2/2+(z_i+t_i)\La^2_i/2\right.\crn
&&\left. \mp \sqrt{\left[\mu^2_2-\mu^2_3+ (\la_4-\la_5)v^2/2+(z_i-t_i)\La^2_i/2\right]^2+\la^2_{10}v^4}\right\}.\eea Given that the $\eta^0$-$\chi^{0*}$ mixing is small, we approximate
 \bea m^2_{H_1}
&\simeq & \mu^2_2+\fr{\la_4}{2}v^2+\fr{z_i}{2} \La^2_i-\fr{\fr 1 4 \la^2_{10}v^4}{\mu^2_{3}-\mu^2_2+\fr 1 2 (\la_5-\la_4)v^2+\fr 1 2 (t_i-z_i)\La^2_i},\\
m^2_{H_2}&\simeq & \mu^2_3+\fr{\la_5}{2}v^2+\fr{t_i}{2} \La^2_i+\fr{\fr 1 4 \la^2_{10}v^4}{\mu^2_{3}-\mu^2_2+\fr 1 2 (\la_5-\la_4)v^2+\fr 1 2 (t_i-z_i)\La^2_i}.\eea
Given in terms of $H_{1,2}$ masses, the mixing angle takes the form, $s_{2\theta}\simeq \la_{10}v^2/(m^2_{H_2}-m^2_{H_1})$, which is similar to that in the toy model.
 
 \subsection{\label{neutrinomass} Neutrino mass}    

\begin{figure}[h]
\bc
\includegraphics[scale=0.9]{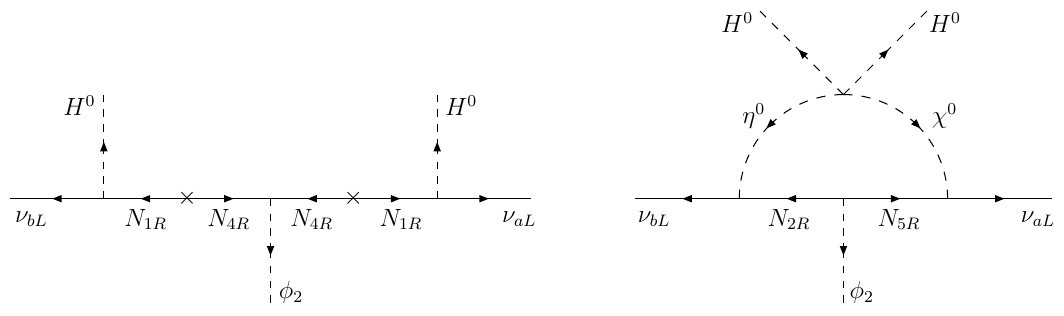}
\caption[]{\label{fig2} Scotoseesaw scheme set by a $U(1)_{B-L}$ gauge completion of $Z_3$.}
\ec
\end{figure}

The diagrams generating neutrino mass are now supplied in Fig. \ref{fig2}, which is a gauge completion of those in Fig. \ref{fig1}. 
 Here, the (tree-level) seesaw part takes into account the contribution of $N_{4R}$ in addition to $N_{1R}$, while the (loop-level) scotogenic part is contributed by the same Dirac fermion $N_{2R,L}$ (where $N_{5R}\equiv N^c_{2L}$ relabels that in the toy model, without confusion), which all appropriately couple to $\phi_2$ fields.  

After the normal scalars develop VEVs, $\langle H\rangle =(0,v/\sqrt{2})$, $\langle \phi_1\rangle = \La_1/\sqrt{2}$, and $\langle \phi_2\rangle =\La_2/\sqrt{2}$, the neutral fermion sector obtains a mass Lagrangian,
\bea \mathcal{L}_{\mathrm{Yukawa}} &\supset& -\fr 1 2 \begin{pmatrix}
\bar{\nu}_{aL} & \bar{N}^c_{1R} & \bar{N}^c_{4R}\end{pmatrix}
\begin{pmatrix}
0 & m_{a1} & 0\\
m_{b1} & M_1 & D_1\\
0 & D_1 & M_4 
\end{pmatrix} 
\begin{pmatrix}
\nu^c_{bL}\\
N_{1R}\\
N_{4R}
\end{pmatrix}\crn
&&- D_2 \bar{N}^c_{2R} N_{5R} - D_3 \bar{N}^c_{3R} N_{6R}+H.c., \eea where $m_{a1}=-h^\nu_{a1}v/\sqrt{2}$, $M_1=-f_1 \La_2/\sqrt{2}$, $M_4=-f_2 \La_2/\sqrt{2}$, $D_2=-f_3\La_2/\sqrt{2}$, and $D_3=-f_4\La_1/\sqrt{2}$. At this stage, it is noted that $M_1\sim M_4\sim D_2\ll m_{a1}\ll D_3$, because of $\La_2\ll v\ll \La_1$. Assuming $D_1\gg m_{a1}\gg M_{1,4}$, the tree-level mass matrix takes the form of inverse seesaw \cite{mohapatraadd3,mohapatraadd4,bernabeuadd5}, as determined by the left diagram in Fig. \ref{fig2}, yielding 
\be m^{\mathrm{tree}}_\nu \simeq \fr{m^2 M_4}{D^2_1}.\label{trdd} \ee Here, note that $M_1$ does not contribute to the neutrino mass. Equivalently, there is a tree-level diagram similar to the toy model in which $N_{1R}N_{1R}$ is now coupled to $\phi_2$, but it gives a negligible contribution, as omitted. In other words, when the gauge completion is given, the inverse seesaw is recognized instead of the canonical seesaw. Indeed, comparing the result with the previous model, we do not need a hierarchy in the Yukawa coupling, namely $h^\nu_{a1}\sim 10^{-2}$ like those of $h^\nu_{a2(3)}$ and $\la_{10}$, given that $M_4/D_1\sim 10^{-3}$, since $D_1$ is at TeV, while $M_4$ is at GeV like the dark matter mass, as expected. It is noted that $N_{1R}$ and $N_{4R}$ now make a pseudo-Dirac state with mass $D_1$, unsuppressed by any symmetry. It is naturally to take $D_1$ at the highest scale of the model, i.e. $D_1\sim \La_1\gg v$, at TeV scale, as desirable. 

The Dirac fermion $N_{2L,R}$ that contributes to radiative neutrino mass takes a mass, namely $D_2=-f_2\La_2/\sqrt{2}$, at $\La_2$ scale, which is radically below the weak scale, $v$. Therefore, the loop correction supplied in the right diagram in Fig. \ref{fig2} is similar to the previous case, for which the neutrino data requires $D_2=-f_2\La_2/\sqrt{2}\sim 10$ GeV, with the choice of relevant parameters similar to the previous model. It is stressed that the role of $\phi_2$ is very important in the neutrino mass generation. Indeed, it is associated with an approximate symmetry so that its VEV, $\La_2$, like $\kappa_3$ generated is small. As a result, both the tree-level (\ref{trdd}) and loop-level (\ref{nmloop}) neutrino masses are proportional to this scale, i.e. $M_4\sim D_2\sim\La_2$, and are doubly suppressed as $(v/\La_1)^2\ll 1$, since $m\sim v$ and $D_1\sim m_{H_{1,2}}\sim \La_1$.       

\subsection{\label{darkmatter} Dark matter} 

The model provides a viable dark matter candidate to be a dark Dirac fermion $N_{2R,L}$ with a mass $D_2$ at GeV regime, implied by the neutrino mass generation (or, in other words, being a consequence of the given benchmark of the parameter space). Here, it is noted that other dark fermions $N_{1R,L}$ (a pseudo-Dirac field) and $N_{3R,L}$ (a pure-Dirac field) have large masses $D_{1,3}$, respectively, at TeV scale. Also, the dark scalars $H_{1,2}$ and $\eta^-,\chi^-$ all have a mass at TeV scale. That said, the dark matter candidate $N_{2R,L}$ is the lightest $Z_3$ field, stabilized by the $Z_3$ conservation.

\begin{figure}[h]
\bc
\includegraphics[scale=1]{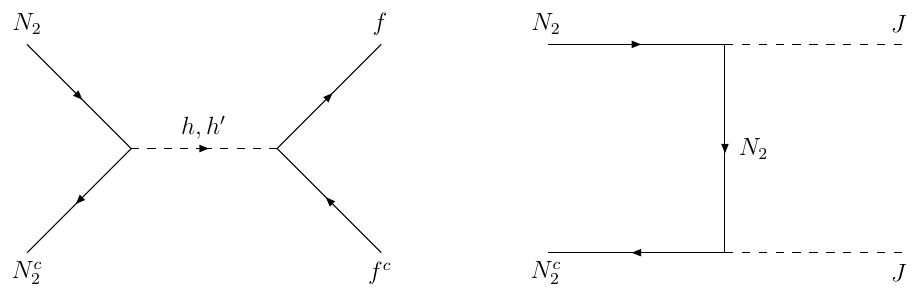}
\caption[]{\label{fig3} Dirac dark matter annihilation set by a $U(1)_{B-L}$ gauge completion of $Z_3$.}
\ec
\end{figure}

In the early universe, $N_{2R,L}$'s annihilate to the standard model particles via $h,h'$ portals due to a $\phi_2$-$H$ mixing. They also annihilate directly to the Majoron $J$'s via $t,u$-channels. The Feynman diagrams that describe the dark matter annihilations are depicted in Fig. \ref{fig3}, where the only $t$-channel diagram is supplied, while the $u$-channel diagram obtained by interchanging the identical $J$ fields is omitted, without confusion.

The annihilation cross section of $N_2$'s to fermions is given by
\be \langle \sigma v_{\mathrm{rel}} \rangle_{N_2N_2\to ff} = \sum_f \fr{N_f s^2_\xi c^2_\xi m^2_f D^4_2}{8\pi v^2 \La^2_2}\left(\fr{1}{4D^2_2-m^2_{h'}}-\fr{1}{4D^2_2-m^2_{h}}\right)^2\left(1-\fr{m^2_f}{D^2_2}\right)^{\fr 3 2}\Theta(D_2-m_f),\label{huytof}\ee where $\Theta(\cdots)$ is the Heaviside step function, and $N_f$ is the color number of $f$. Additionally, the annihilation cross section of $N_2$'s to Majorons is computed by
\be \langle \sigma v_{\mathrm{rel}} \rangle_{N_2N_2\to JJ} =\fr{\langle v^2\rangle D^6_2 (D^2_2-m^2_J)^2}{6\pi \La^4_2(2D^2_2-m^2_J)^4}\left(1-\fr{m^2_J}{D^2_2}\right)^{\fr 1 2}\Theta(D_2-m_J),\label{huytoj} \ee where $\langle v^2\rangle =3/2x_F$ for $x_F=D_2/T_F\simeq 25$ at freeze-out temperature, and $v=\fr 1 2 v_{\mathrm{rel}}$ is dark matter velocity, without confusion with the weak scale.

\begin{figure}[h]
\bc
\includegraphics[scale=0.55]{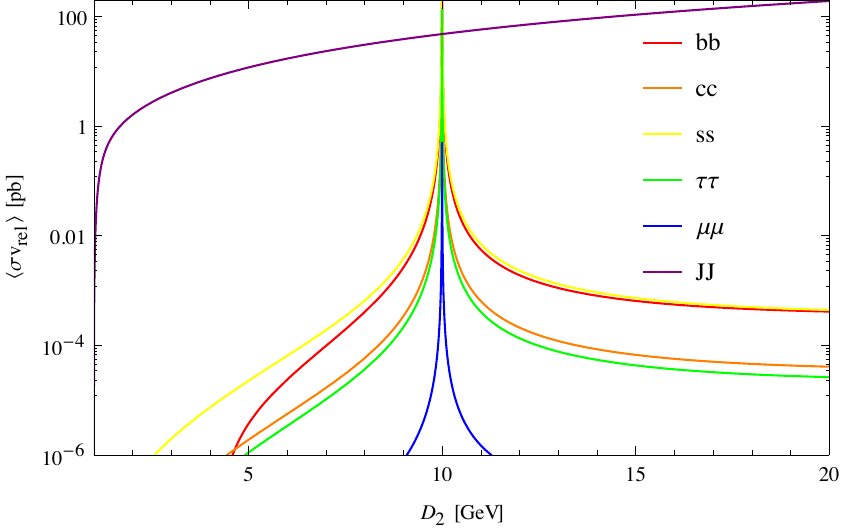}
\includegraphics[scale=0.55]{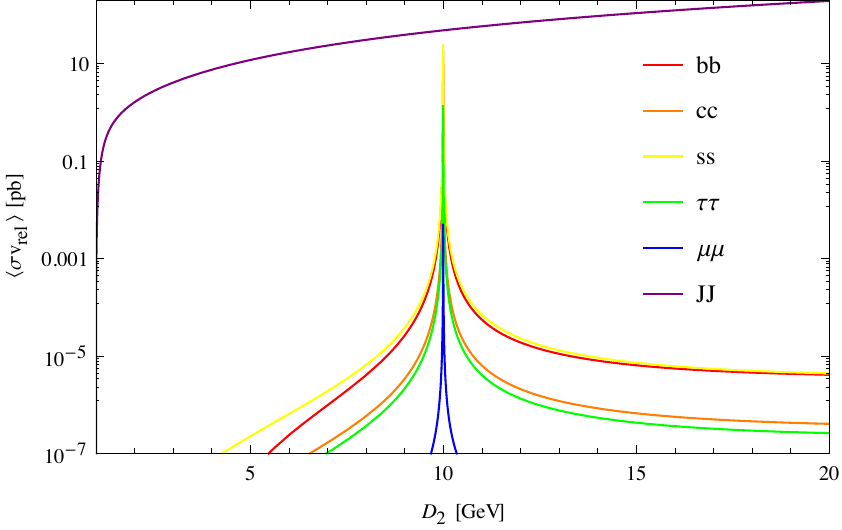}
\caption[]{\label{fig4} Dark matter annihilation cross sections to various channels plotted as functions of the dark matter mass for $\xi=10^{-2}$ (left panel) and $10^{-3}$ (right panel) for a comparison.}
\ec
\end{figure}   

To give a numerical investigation, we use $m_s=0.093$, $m_c=1.273$, $m_b=4.183$, $m_\mu=0.105$, $m_\tau=1.776$, $m_h=125$, and $v=246$, which are all in GeV \cite{pdg}. Additionally, we take $\La_2=20$ GeV and $m_{h'}=20$ GeV. In Fig. \ref{fig4}, we plot the dark matter annihilation cross sections to various fermions and Majorons as functions of the dark matter mass, $D_2$, for a comparison, where $m_J=1$ GeV is taken. It is clear that the annihilation to Majorons as in (\ref{huytoj}) is independent of the Higgs mixing, $\xi$, whereas the annihilation to fermions as in (\ref{huytof}) is quite sensitive to this mixing. The left and right panels correspond to the choice of $\xi=10^{-2}$ and $10^{-3}$ for demonstration, respectively. 

\begin{figure}[h]
\bc
\includegraphics[scale=1]{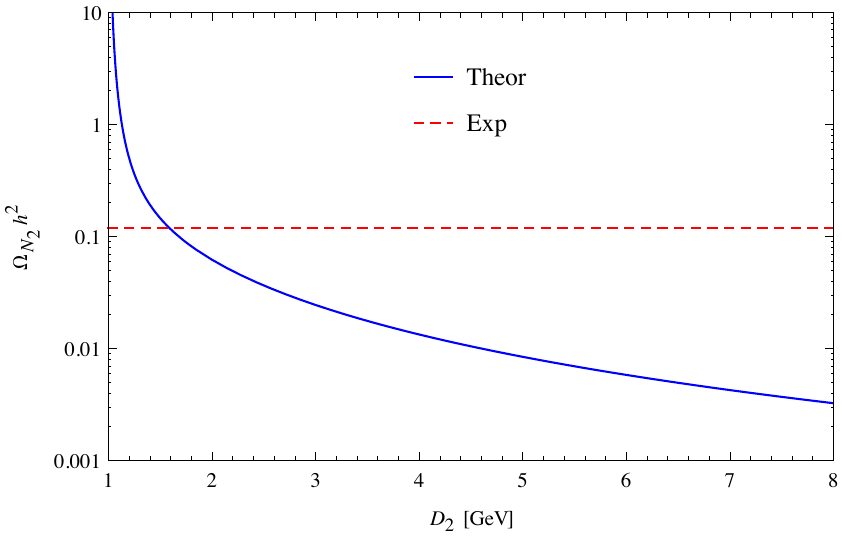}
\caption[]{\label{fig5} Dark matter relic density set by its annihilation to the Majoron.}
\ec
\end{figure}   

It is clear that if the dark matter is heavier than $J$, the dark matter annihilation to $J$ dominates and sets the relic density independent of $\xi$, as depicted in Fig. \ref{fig5}. The correct abundance requires the dark matter mass $D_2>1.6$ GeV.

\begin{figure}[h]
\bc
\includegraphics[scale=0.55]{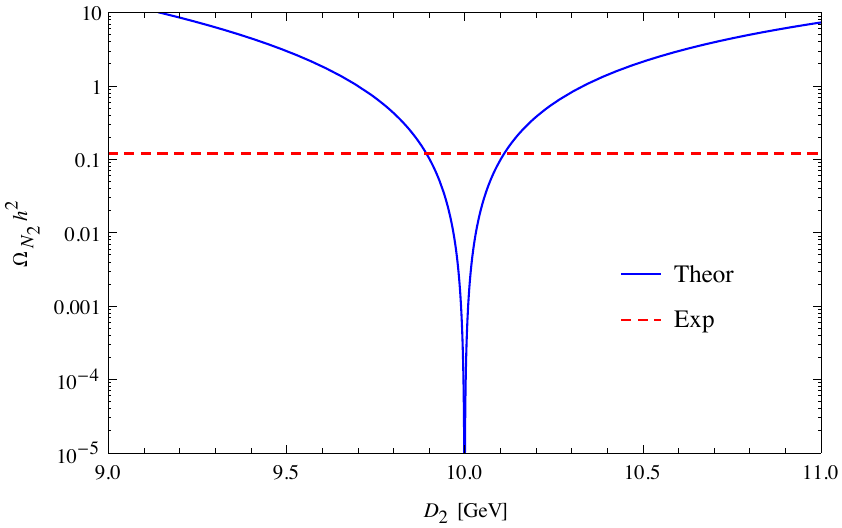}
\includegraphics[scale=0.55]{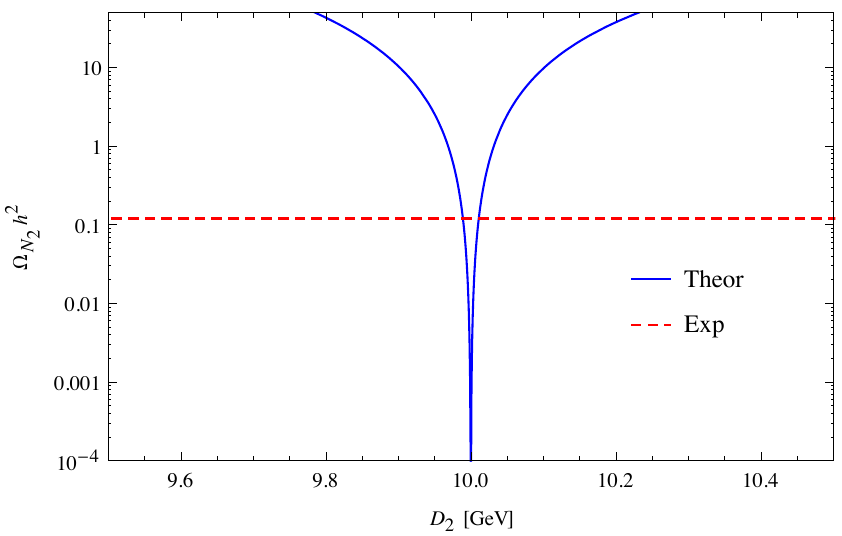}
\caption[]{\label{fig6} Dark matter relic density governed its annihilation to fermions by the $h'$ resonance.}
\ec
\end{figure}   

Otherwise, if the dark matter is lighter than $J$, the annihilation to $J$ is suppressed. In this case, the $h'$ resonance governs the relic density, as given in Fig. \ref{fig6}, according to $\xi=10^{-2}$ (left panel) and $10^{-3}$ (right panel). The viable regime, i.e. the width of the resonance, depends on the Higgs mixing parameter. The correct abundance acquires the dark matter mass to be $9.89\ \mathrm{GeV} <D_2<10.11$ GeV and $9.98\ \mathrm{GeV} <D_2<10.01$ GeV corresponding to $\xi=10^{-2}$ and $10^{-3}$, respectively.

In the direct detection, the dark matter $N_2$ scatters with a nucleon ($N=p,n$) through $h,h'$ exchanges, which couple to both $N_2$ and quarks confined in the nucleon. This process is described by the effective interaction at quark level as \cite{dd1}
\be \mathcal{L}_{\mathrm{eff}}\supset \al^S_q (\bar{N}_2 N_2)(\bar{q}q),\ee where the effective coupling is derived by \be \al^S_q = -\fr{s_\xi c_\xi m_q D_2}{v\La_2}\left(\fr{1}{m^2_h}-\fr{1}{m^2_{h'}}\right).\ee 

The scattering cross section of dark matter with nucleon is summed over quark contributions multiplied by relevant nucleon form factors, such as 
\be \sigma^{\mathrm{SI}}_{N_2\mathrm{-}N}=\fr{4m^2_r}{\pi}(f^N)^2,\ee where $m_r=m_N D_2 /(m_N+D_2)$ and \be \fr{f^N}{m_N}=\sum_{q=u,d,s}\fr{\al^S_q}{m_q}f^N_{Tq}+\fr{2}{27}f^N_{TG}\sum_{q=c,b,t}\fr{\al^S_q}{m_q}\simeq -\fr{0.35 s_\xi c_\xi D_2}{v\La_2}\left(\fr{1}{m^2_{h}}-\fr{1}{m^2_{h'}}\right),\ee where the form factors obey $f^N_{TG}=1-\sum_{q=u,d,s}f^N_{Tq}$, $f^p_{Tu}=0.02$, $f^p_{Td}=0.026$, $f^p_{Ts}=0.118$, $f^n_{Tu}=f^p_{Td}$, $f^n_{Td}=f^p_{Tu}$, and $f^n_{Ts}=f^p_{Ts}$ \cite{dd2}. 

Hence, we compare the direct detection cross section according to the case of the resonance dark matter annihilation, 
\bea \sigma^{\mathrm{SI}}_{N_2\mathrm{-}N} &\simeq& \fr{0.49m^2_r m^2_N s^2_\xi c^2_\xi D^2_2}{\pi v^2 \La^2_2}\left(\fr{1}{m^2_h}-\fr{1}{m^2_{h'}}\right)^2\crn
&\simeq& \left(\fr{s_\xi}{10^{-2}}\right)^2\left(\fr{D_2}{10\ \mathrm{GeV}}\right)^2\times 10^{-43}\ \mathrm{cm}^2\\
&\simeq& \left(\fr{s_\xi}{10^{-3}}\right)^2\left(\fr{D_2}{10\ \mathrm{GeV}}\right)^2\times 10^{-45}\ \mathrm{cm}^2, \eea where we take $m_N=1$ GeV, $m_{h'}=20$ GeV, and $\La_2=20$ GeV, as before. The prediction with $\xi\sim 10^{-2}$ is large, already ruled out by the experimental limit for the resonance dark matter annihilation with a mass in the regime $D_2\sim 10$ GeV \cite{dd3}. However, the prediction with $\xi\sim 10^{-3}$ is in agreement with the experiment $\sigma^{\mathrm{SI}}_{\mathrm{exp}}\sim 10^{-45}\ \mathrm{cm}^2$. In this case, the viable regime of dark matter mass is very narrow, as given above.

In the alternative case of the dark matter annihilation to Majorons, the dark matter annihilation cross section to Majorons is independent of $\xi$, as given, which does not give any restrict on $\xi$. The direct detection cross section computed in this benchmark is  
\bea \sigma^{\mathrm{SI}}_{N_2\mathrm{-}N} 
&\simeq& \left(\fr{s_\xi}{10^{-2}}\right)^2\left(\fr{D_2}{1.6\ \mathrm{GeV}}\right)^2\times 3\times 10^{-45}\ \mathrm{cm}^2.\eea Although the existing direct detection experiment has not given any constraint on a WIMP with mass below $9$ GeV \cite{dd3}, our model predicts $\sigma^{\mathrm{SI}}_{N_2\mathrm{-}N}\sim 3\times 10^{-45}\ \mathrm{cm}^2$ for the dark matter mass $D_2=1.6$ GeV and the Higgs mixing angle $\xi\sim 10^{-2}$ as bounded by the Higgs measurement. It is worth exploring in the shortcoming experiment.              

\subsection{Charged lepton flavor violation}

We assume that the charged leptons $e_a$ ($a=1,2,3$) are flavor diagonal, without loss of generality. That said, $e_a=e,\mu,\tau$ for $a=1,2,3$, respectively, are physical fields by themselves. The Yukawa interactions of $\eta,\chi$ in (\ref{adt3001a}) lead to charged lepton flavor violation processes, e.g. $e_a\to e_b\gamma$ and $e_{a}\to 3e_b$ if kinematically allowed and $\mu$-$e$ conversion in nuclei, at one-loop levels. The relevant couplings are 
\bea \mathcal{L}_{\mathrm{Yukawa}} &\supset& h^\nu_{a2}\bar{l}_{aL}\eta N_{2R}+h^\nu_{a3}\bar{l}_{aL}\chi N_{5R}+H.c.\crn
&\supset& h^\nu_{a2}\bar{e}_{aL}\eta^- N_{2R}+h^\nu_{a3}\bar{e}_{aL}\chi^- N^c_{2L}+H.c.,\eea where $N_{5R}\equiv N^c_{2L}$ such that $N_2=N_{2L}+N_{2R}$ is a physical pure-Dirac field, as well as noticing that $\eta^-$ and $\chi^-$ are physical charged scalar fields, as aforementioned.

\begin{figure}[h]
\bc
\includegraphics[scale=1]{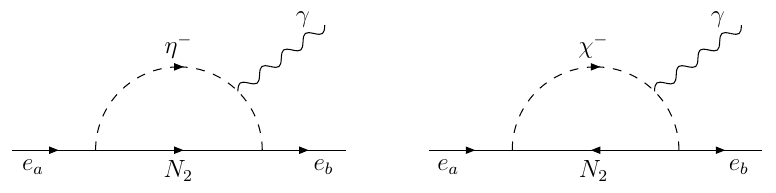}
\caption[]{\label{fig7} Charged lepton flavor violation process $e_a\to e_b\gamma$ proceeded through the mediators of dark fields, say charged scalars $\eta^-,\chi^-$ and neutral Dirac field $N_{2}$.}
\ec
\end{figure}

In what follows, we consider only $e_a\to e_b\gamma$ since if such process is constrained, the remaining processes are easily evaded (see, e.g., \cite{chekkaladd6}). The concerning process occurs via Feynman diagrams contributed by dark fields $\eta^-,\chi^-,N_2$ as depicted in Fig. \ref{fig7}.
It is straightforward to compute the branching decay ratio,
\be \mathrm{Br}(e_a\to e_b\gamma)=\fr{3\al v^4}{32\pi}\left|\fr{h^{\nu *}_{a2}h^\nu_{b2}}{m^2_{\eta^-}}F\left(\fr{D^2_2}{m^2_{\eta^-}}\right)+\fr{h^{\nu *}_{a3}h^\nu_{b3}}{m^2_{\chi^-}}F\left(\fr{D^2_2}{m^2_{\chi^-}}\right)\right|^2\mathrm{Br}(e_a\to e_b \nu_a \nu^c_b),\ee where the loop function $F(x)=(1-6x+3x^2+2x^3-6x^2\ln x)/6(1-x)^4$ possesses no pole at $x=1$ and decreases when $x$ increases from zero, thus $F(x)<1/6$ for any $x>0$. This result for the pure-Dirac mediator is in agreement to, i.e. identical to, that for the pure-Majorana mediator \cite{tomaadd7}, which is not surprised. 

Let us evaluate the most severe process $\mu\to e\gamma$, such as 
\be \mathrm{Br}(\mu\to e\gamma)\leq 0.95\times 10^{-15}\left(\fr{|h^{\nu*}_{\mu k} h^\nu_{e k}|}{10^{-4}}\right)^2\left(\fr{1\ \mathrm{TeV}}{m_{\eta^-,\chi^-}}\right)^4,\ee with the aid of $m_{\eta^-}\sim m_{\chi^-}$, $h^\nu_{a2}\sim h^\nu_{a3}$, $\al=1/128$, $v=246$ GeV, and $\mathrm{Br}(\mu\to e \nu_\mu \nu^c_e)\simeq 1$, where $k=2,3$. This rate is below the current bound $\mathrm{Br}(\mu\to e\gamma)\simeq 4.2\times 10^{-13}$ \cite{baldiniadd8} and even the projected sensitivity $\mathrm{Br}(\mu\to e\gamma)\simeq 6\times 10^{-14}$ \cite{baldiniadd9}, for the given benchmark $h^\nu_{ak}\sim 10^{-2}$ and $m_{\eta^-,\chi^-}\sim 1$ TeV. In other words, the charged lepton flavor violation in this setup is negligible, compared with the existing limits.

Last, but not least, there might exist a Yukawa coupling of the heavy field $N_{6R}$ with usual leptons and $\eta$, say $h'\bar{l}_{L}\eta N_{6R} +H.c.$, which does not contribute to neutrino mass and dark matter, as already mentioned. However, it might give rise to the charged lepton flavor violation at one-loop level as mediated by the dark charged scalar $\eta^-$ and the dark pure-Dirac fermion $N_3=N_{3L}+N_{3R}$, where $N_{3L}\equiv N^c_{6R}$. This contribution is easily evaded by taking $h'$ to be small, since $h'$ is not constrained by the neutrino mass and dark matter.           
        
\section{\label{concl} Conclusion}

We have argued that the center of the color group sets the exotic right-handed neutrino sector in such a way that the neutrino mass and the dark matter candidate obey a scotoseesaw mechanism. This idea has been realized in a gauge completion with $U(1)_{B-L}$ such that it conserves a residual group $Z_3=\{1,T,T^2\}$ with $T=w^{3(B-L)}$ isomorphic to the center of the color group. That said, the seesaw part works properly in terms of an inverse seesaw, while the scotogenic part naturally implies a light Dirac dark matter, which are all set by a second singlet scalar associated with an approximate extra symmetry, alternative to $B-L$. Two scenarios for the Dirac dark matter are recognized, (i) It can obtain a mass at 10 GeV and a relic density through the $h'$ resonance annihilation, but the direct detection restricts the $h$-$h'$ mixing to be $\xi\sim 10^{-3}$ and (ii) It can obtain a mass at 1.6 GeV and a relic density through the annihilation to Majoron $J$'s, predicting a desirable direct detection cross section at $3\times 10^{-45}\ \mathrm{cm}^2$ for $\xi\sim 10^{-2}$ appropriate to the standard model Higgs measurement.  

\section*{Ackowledgements}    

This research is funded by Vietnam National Foundation for Science and Technology Development (NAFOSTED) under grant number 103.01-2023.50.

\end{document}